\documentclass[11pt]{article}
\usepackage{txfonts} 
\usepackage{cospar}
\usepackage[sectionbib]{natbib}
\renewcommand{\bibsection}{\section*{REFERENCES}}
\renewcommand{\bibhang}{\setlength{8mm}}
\renewcommand{\bibsep}{\setlength{0mm}}
\usepackage{graphicx}
\usepackage[figuresright]{rotating}

\title{DO WE REALLY OBSERVE A BOW SHOCK IN N157B...?}

\author{E. van der Swaluw\address{Dublin Institute for Advanced Studies, 5 Merrion Square,
        Dublin 2, Ireland}}

\begin{document}

\maketitle

\begin{abstract}
I present a model of a pulsar wind interacting with its associated supernova remnant. I will use
the model to argue that one can explain the morphology of the pulsar wind nebula inside N157B,
a supernova remnant in the Large Magellanic Cloud,
without the need for a bow shock interpretation. The model uses a hydrodynamics code which 
simulates the evolution of a pulsar wind nebula, when the pulsar is moving at a high velocity 
($1\; 000$ km/sec) through the expanding supernova remnant.
The evolution of the pulsar wind nebula can roughly be divided into three stages. In the
first stage the pulsar wind nebula is expanding supersonically through the freely expanding ejecta
of the progenitor star ($\sim 1\; 000$ years). In the next stage the expansion of the pulsar wind nebula is not
steady, due to the interaction with the reverse shock of the supernova remnant; the pulsar wind nebula
oscillates violently between contraction and expansion, but will ultimately relax towards a steady subsonic
expansion ($\sim 1\; 000 - 10\; 000$ years). The last stage occurs when the head of the pulsar wind 
nebula, containing the active pulsar, deforms into a bow shock ($> 10\; 000$ years), due to the motion 
of the pulsar becoming supersonic. Ultimately it is this bow shock structure bounding the pulsar, which 
directly interacts with the shell structure of the supernova remnant, just before the pulsar breaks out of the 
supernova remnant. I will argue that the pulsar wind nebula inside N157B is currently in the second
stage of its evolution, i.e. the expansion of the pulsar wind nebula is subsonic and there is no bow shock 
around the pulsar wind bubble. The strongly off-centered position of the pulsar with respect to its
pulsar wind nebula is naturally explained by the result of the interaction of the reverse shock with the 
pulsar wind nebula, as the simulation bears out.
\end{abstract}

\section*{A PULSAR WIND INSIDE A SUPERNOVA REMNANT}

A supernova explosion marks the end of the evolutionary track of a massive star, but
launches the beginning of the evolution of a supernova remnant (SNR). If the fossil of the progenitor
star is an active pulsar, the associated pulsar wind drives a pulsar wind nebula (PWN) into
the surrounding medium (Rees and Gunn 1974). The evolution of the PWN is coupled and determined by
the evolution of the SNR, because the total energy release of the relativistic pulsar wind over the
pulsar's lifetime is
small ($10^{49}-10^{50}$ erg) compared with the total mechanical energy of the SNR ($10^{51}$ erg).

A young SNR ($\sim 100-1\; 000$ years) consists of a twofold shock structure, i.e. a 
forward shock which propagates into the interstellar medium (ISM) and a reverse shock which forms
due to the deceleration of the forward shock by the surrounding ISM (McKee 1974). When the forward 
shock has swept up a comparable amount of mass as was ejected in the explosion event, the reverse shock 
propagates back into the interior of the SNR, where the PWN is expanding through the freely expanding 
ejecta of the progenitor star. As the reverse shock encounters the PWN, the latter is crushed, due 
to the huge pressure downstream of the reverse shock relative to the pressure inside the PWN 
(van der Swaluw et al. (2001); Blondin et al. (2001)).

Van der Swaluw et al. (2001) have shown that in the case of a centered pulsar, the interaction between 
the reverse shock and the centrally powered PWN results in an unsteady expansion stage, during which the interior 
of the pulsar wind bubble adjusts itself to the pressure of the surrounding SNR. Ultimately the 
expansion of the PWN becomes subsonic. However, when the pulsar is born with a kick velocity, the position 
of the active pulsar has become off-center with respect to the SNR at this stage, and will ultimately overtake 
and break through the shell of the decelerating SNR. Before this break-through event the motion of the pulsar
becomes supersonic which deforms the head of the PWN, containing the active pulsar, into a bow shock 
(van der Swaluw et al. 1998).

In this paper I present results from a hydrodynamical simulation, carried out with the Versatile Advection
Code (VAC)\footnote{See http://www.phys.uu.nl/\~{}toth/}. The simulation describes the evolution 
of a pulsar wind nebula when the pulsar is moving at a high velocity ($1\; 000$ km/sec) through its 
associated supernova remnant. The simulation shows that the position of the pulsar is strongly off-centered
with respect to its pulsar wind bubble, after the interaction between the reverse shock and the PWN.
Furthermore the simulation confirms the formation of a bow shock at a later stage of the PWN, as
the motion of the pulsar becomes supersonic.
The pulsar inside the SNR N157B is strongly off-centered with respect to its PWN. However the age of the
remnant is rather young to explain the PWN morphology as a bow shock. Therefore I propose a scenario
for which the pulsar wind nebula inside N157B has been recently crushed by the reverse shock. The simulation
shows that this explains the off-centered position of the pulsar with respect to its PWN, the age of the remnant 
and the overall morphology of the PWN in a natural way. 

\section*{IMPORTANT TIMESCALES FOR PWN EVOLUTION}
  
Initially the PWN is bounded by a strong shock propagating into the freely expanding ejecta of
the progenitor star. As the forward shock of the SNR is expanding into the ISM, it starts to 
decelerate. McKee and Truelove (1995) define a timescale
$t_{\rm ST}$, which marks the age of the remnant when it has swept up roughly 1.61 times the ejected
mass $M_{\rm ej}$. They show that when the age of the remnant $t_{\rm snr}$, approximately equals 
$t_{\rm snr}\simeq 5 \; t_{\rm ST}$, the reverse shock hits the center of the SNR. Therefore one can 
roughly equal this timescale with the age at which the PWN interacts with the reverse shock:
\begin{equation}
t_{\rm rev}\;\simeq\; 5t_{\rm ST}\; =\;  1\; 045 E^{-1/2}_{51}
\left({M_{\rm ej}\over M_\odot}\right)^{5/6}n_0^{-1/3}\;\;{\rm years}\; ,
\end{equation}
here $E_{51}$ is the total mechanical energy of the SNR in units of $10^{51}$ erg 
and $n_0$ is the ambient hydrogen number density assuming an interstellar composition
of 10 H : 1 He. The above timescale is very close to the one given by Reynolds and Chevalier 
(1984) and marks the end of the stage of the PWN during which it is expanding supersonically. 

After the interaction of the PWN with the reverse shock, the pulsar is propagating through a 
medium which has been reheated due to the passage of the reverse shock. Therefore the motion of
the pulsar and the expansion of the PWN are both subsonic. Van der Swaluw et al. (1998) have 
shown that the motion of the pulsar becomes supersonic and deforms the pulsar wind bubble 
into a bow shock, when the position of the pulsar $R_{\rm psr}$, with respect to the radius of 
the SNR shock $R_{\rm snr}$, equals: 
\begin{equation}
R_{\rm psr}\simeq 0.677 R_{\rm snr},
\end{equation}
assuming a Sedov-Taylor expansion rate for the SNR. This occurs at roughly half the
crossing time, i.e. the age of the SNR when the pulsar overtakes the shell of the remnant:
\begin{equation}
\label{crosstime}
       t_{\rm cr}\;\simeq\; 1.4 \times 10^4 \; E_{51}^{1/3} V_{\rm 1000}^{-5/3}n_0^{-1/3}
       {\rm years} \; ,
\end{equation}       
here $V_{1000}$ is the velocity of the pulsar in units of $1\; 000$ km/sec. Using the above timescales,
one can roughly summarise the PWN evolution inside a Sedov-Taylor SNR as:

\noindent $\bullet$ a supersonically expansion stage ($t_{\rm snr} < t_{\rm rev}$)

\noindent $\bullet$ a subsonically expansion stage ($t_{\rm rev} < t_{\rm snr} < 0.5 t_{\rm cr}$)

\noindent $\bullet$ a bow shock stage ($0.5 t_{\rm cr} < t_{\rm snr} < t_{\rm cr}$)

\section*{HYDRODYNAMICAL SIMULATIONS}

\subsection*{Simulation Method}

I performed a hydrodynamical simulation of a SNR with a mechanical energy of $E_0=10^{51}$ 
erg and a total ejecta mass of $M_{\rm ej}=3M_\odot$ expanding into a uniform ISM with a 
hydrogen number density of $n_0=0.43$. The simulation is performed in the restframe of the
pulsar (wind), placed at the (explosion) center of the remnant with a constant luminosity 
$L_{\rm pw}=10^{38}$ ergs/sec and a pulsar velocity 
$V_{\rm psr}=1\; 000$ km/sec. The pulsar wind luminosity is taken to be constant throughout the simulation.
This is required in order to resolve the pulsar wind termination shock. The results are certainly not changed
qualitatively, because the total integrated energy input by the pulsar wind is only a $\sim 5$ \% of
the total mechanical energy of the SNR at the end of the simulation. This insures that the evolution of 
the pulsar wind bubble is completely determined by its surrounding evolving SNR. 
I used the Versatile Advection Code (T\'oth 1996) to solve the equations
of gas dynamics on a uniform grid with axial symmetry, using a cylindrical coordinate system
$(R,z)$ in the restframe of the pulsar. I simulate the pulsar wind by depositing mass $\dot M$ and
energy $L_{\rm pw}$ continuously in a few grid cells concentrated around the position of the pulsar. The terminal
velocity of the pulsar wind is determined from these two parameters, i.e. $v_{\infty}=\sqrt{2L_{\rm pw}/\dot M}$,
and has a value much larger then all the other velocities of interest in the 
simulation. The current version of the VAC code does not include relativistic hydrodynamics, therefore
the best approach available is to take an adiabatic index of the relativistic fluid $\gamma_{\rm pwn}=5/3$. 

\begin{minipage}{150mm}
\includegraphics[width=150mm]{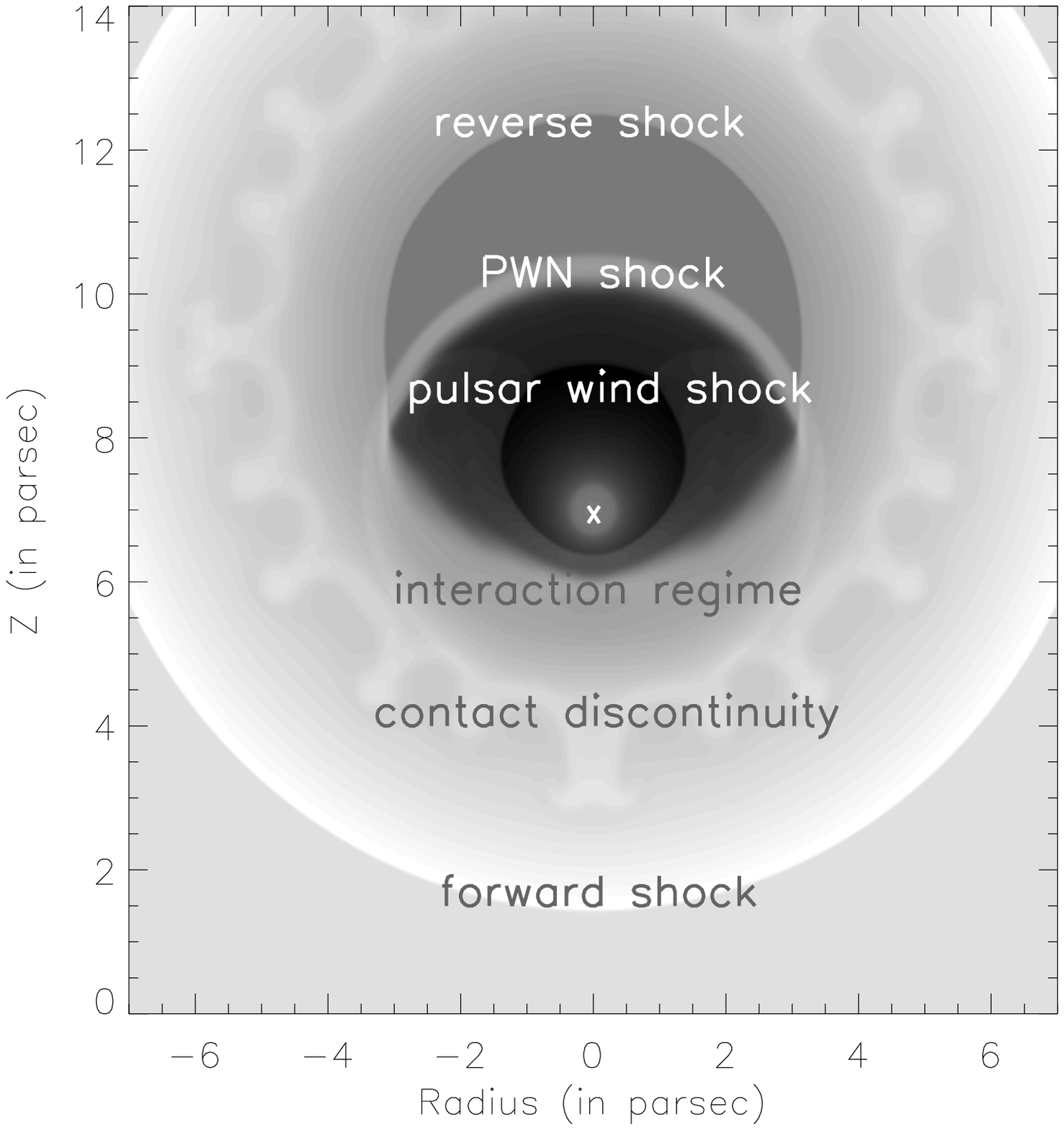}


\centerline{Fig. 1. Logarithmic gray-scale representation of the density distribution at an age 
$t_{\rm snr}=2\; 250$ years.}

\end{minipage}
\bigskip
\subsection*{The Supersonically Expansion Stage (Figure 1)}

Figure 1 shows the density profile of the PWN/SNR system at an age of 
$t_{\rm snr}= 2\; 250$ years. The PWN has been dragged towards the reverse shock by the motion
of the pulsar. The front part of the PWN has already been crushed by the reverse shock (the part
indicated in the figure by {\it interaction regime}), it will 
take another $\sim 500$ years before the whole PWN will be completely crushed by the reverse shock. 
One can clearly distinguish the forward shock, contact discontinuity, PWN shock and the pulsar wind 
(termination) shock. In the figure the cross corresponds with the pulsar position.

\begin{minipage}{150mm}
\includegraphics[width=150mm]{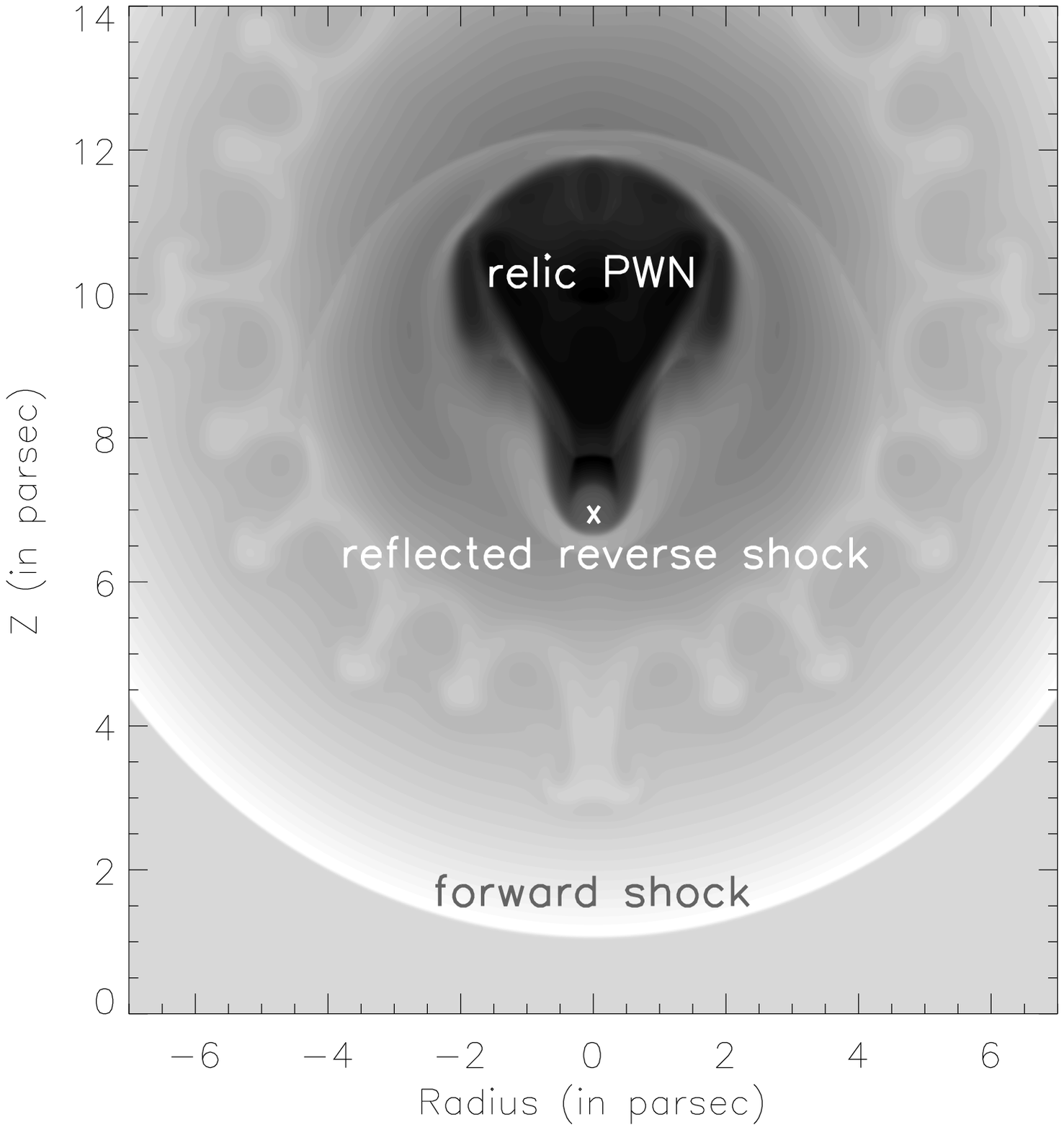}


\centerline{Fig. 2. Logarithmic gray-scale representation of the density distribution at an age 
$t_{\rm snr}=3\; 000$ years. }

\end{minipage}
\bigskip
\subsection*{The subsonically Expansion Stage (Figure 2)}

Figure 2 shows the density profile of the PWN/SNR system at an age of $t_{\rm snr}= 3\; 000$ years, at 
which time the whole PWN has been crushed by the reverse shock. One can clearly distinguish the 
crushed {\it relic} part of the bubble blown in the initial stage of the PWN, when it was bounded by
a strong shock. The pulsar is positioned at the head of the PWN (indicated by a cross again), {\it but} 
the feature ahead of the pulsar is {\it not} a bow shock but is the reflected reverse shock. 
{\it Notice that although there is no bow shock around the PWN, the pulsar is strongly off-centered 
with respect to the PWN, after its interaction with the reverse shock.}

\subsection*{The bow shock stage (Figure 3)}

Figure 3 shows the density profile of the PWN/SNR system at an age of $t_{\rm snr}= 15\; 750$ years,
at the end of the simulation. The pulsar is approaching the shell of the SNR and the head of the pulsar wind
bubble, containing the active pulsar, has been deformed into a bow shock.The simulation confirms 
the formation of a bow shock at the head of the PWN, at half the 
crossing time $t_{\rm cr}$, when the position of the pulsar $R_{\rm psr}$ with respect to the shell of 
the SNR $R_{\rm snr}$ equals $R_{\rm psr}\simeq 0.677 R_{\rm snr}$, as predicted by 
van der Swaluw et al. (1998). 
Notice that the lengthscale of the PWN is small compared
with the SNR shock structure. This confirms the approximation made by van der Swaluw et al. (2003) to
model the break-through event of a pulsar wind through the shell of its SNR, by neglecting the
curvature of the SNR blast wave. 

\begin{minipage}{150mm}
\includegraphics[width=150mm]{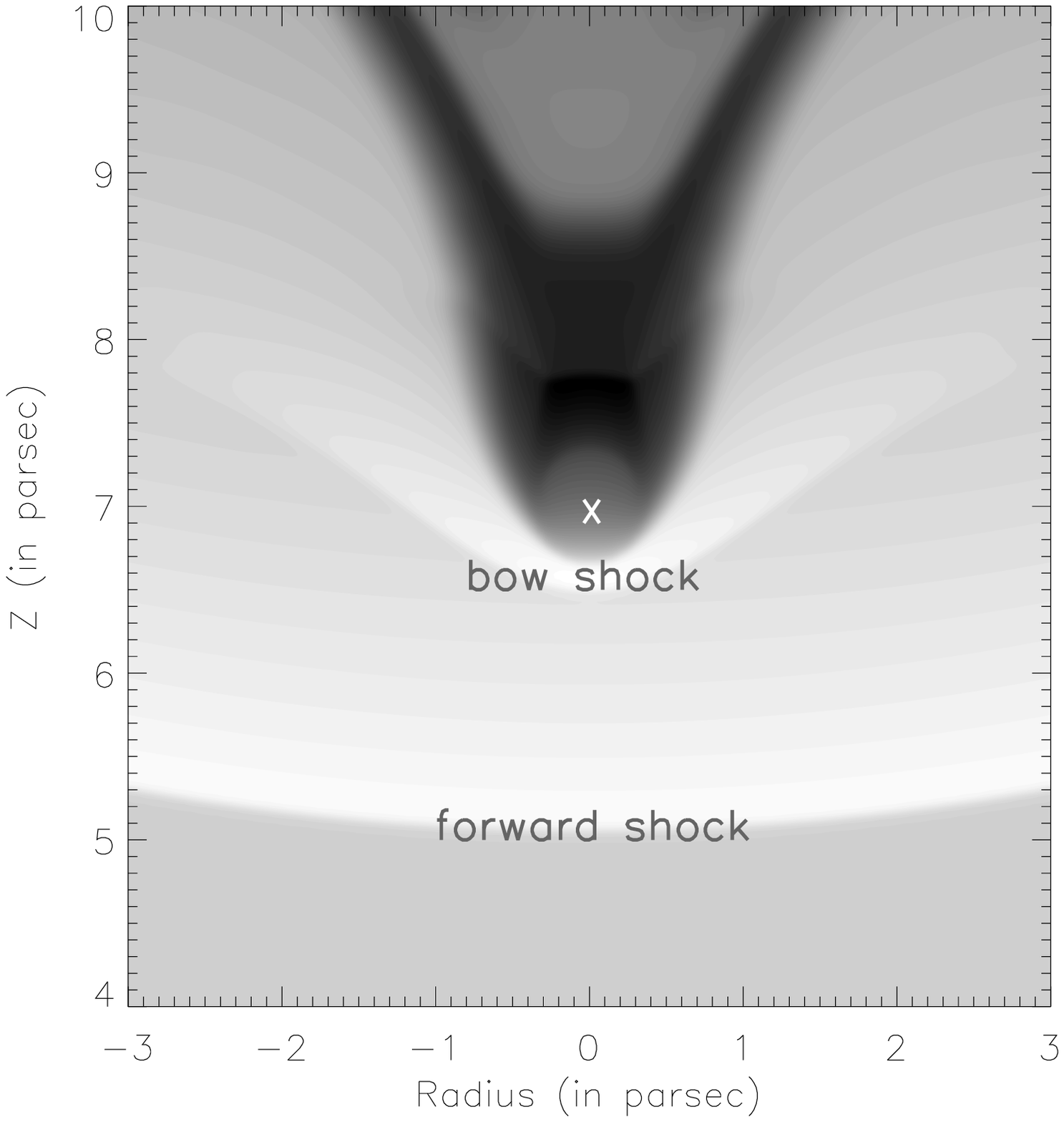}


\centerline{Fig. 3. Logarithmic gray-scale representation of the density distribution at an age 
$t_{\rm snr}=15\; 750$ years.}

\end{minipage}
\bigskip
\subsection*{Summary}

The results of the simulation performed in this paper clearly show the three stages of an
evolving PWN in an expanding SNR; a supersonically expansion stage, a subsonically expansion
stage and a bow shock stage. The simulation bears out that after the interaction between the
reverse shock and the PWN, the pulsar is strongly off-centered with respect to its PWN, while
the motion of the pulsar is still subsonic. Furthermore the simulation shows the formation of
a bow shock at the head of the PWN, containing the active pulsar, at a later stage, when the
motion of the pulsar becomes supersonic. However, the expansion of the relic part of the PWN, 
blown in the supersonically expansion stage, remains subsonic.


\section*{THE PULSAR WIND NEBULA INSIDE SNR N157B}

The bright X-ray object N157B is a young SNR dominated by plerionic emission 
originating from the PWN inside this object (Dickel and Wang 2003). The age 
of the remnant
is approximately $5\; 000$ years old (Wang and Gotthelf 1998) and contains a 16 ms pulsar
(Marshall et al. 1998). The velocity of this pulsar is high 
($V_{\rm psr}\simeq 1\; 000$ km/sec), if one assumes that the pulsar was born in the
central region of the bright radio emission (Lazendic et al. 2000). Wang and Gotthelf 
(1998) argue for a bow shock interpretation of the PWN in N157B. 
A classical example of a pulsar wind bow shock inside its associated SNR is the remnant CTB80,
with an estimated age of $t_{\rm snr}\simeq 100\; 000$ years and an associated pulsar with a spindown luminosity
of $L_{\rm pw}\simeq 4.0\times 10^{36}$ erg/sec. These values show a remarkable contrast with the parameters
of the SNR N157B ($t_{\rm snr}\simeq 5\; 000$ years and $L_{\rm pw}\simeq 4.8\times 10^{38}$ erg/sec).
Furthermore from Figure 2 of Wang et al. (2001) it seems like the position of the pulsar is more 
or less centered in the SNR. This is in contrast with the analysis performed by van der Swaluw et al. (1998), 
which predicts $R_{\rm psr}/R_{\rm snr} \ge 0.677$ for a bow shock PWN, a result confirmed by the simulation in this paper.

In order to solve the issues mentioned above, I propose the following evolutionary stage for
the PWN inside SNR N157B: the PWN has recently been crushed by the reverse shock of its associated
SNR. Due to the combination of the reverse shock interaction and the high velocity of the pulsar, 
the position of the pulsar is strongly off-centered with respect to its PWN, as shown by the results
of the simulations performed in this paper. Figure 2 represents the current morphology of N157B, where 
the relic PWN represents the central bright parts of the radio and X-ray emission in N157B. This 
scenario automatically implies the absence of a bow shock in N157B. Furthermore the age of N157B is in 
rough agreement with this scenario ($t_{\rm rev} < t_{\rm N157B}<0.5 t_{\rm cr}$).


\section*{CONCLUSION}

I have performed a hydrodynamical simulation of a PWN, when the pulsar is
moving at a high velocity through its associated SNR. The simulation shows that one can roughly 
distinguish between a supersonically expanding PWN, a subsonically expanding PWN and a stage 
during which part of the PWN is bounded by a bow shock, due to the supersonic motion of the 
pulsar. After the interaction between the PWN and the reverse shock the position of the pulsar
has become off-centered with respect to its PWN. Therefore one should be careful in distinguishing
a {\it young} subsonically expanding PWN ($t_{\rm rev} < t_{\rm pwn} < 0.5 t_{\rm cr}$) and an
{\it older} PWN with a bow shock around its pulsar ($0.5 t_{\rm cr} < t_{\rm pwn} < t_{\rm cr}$). In the case
of N157B, I have argued for a {\it no} as an answer to the title of this paper.

\vskip 0.2 true cm
\noindent
E-mail address of E. van der Swaluw \hskip 0.1 true cm  swaluw@rijnh.nl

\end{document}